\documentstyle[prl,aps,twocolumn,epsf]{revtex}
\begin{document}
\draft
\title{Multilevel blocking approach to the fermion sign problem 
in path-integral Monte Carlo simulations}
\author{C.H. Mak,$^1$ R.~Egger,$^2$ and H.~Weber-Gottschick$^3$}
\address{${}^1$Department of Chemistry, 
University of Southern California, Los Angeles, CA 90089-0482\\
${}^2$Fakult\"at f\"ur Physik, Albert-Ludwigs-Universit\"at,  
 D-79104 Freiburg, Germany\\
${}^3$Institut f\"ur Theoretische Physik, Universit\"at Stuttgart,   
 D-70550 Stuttgart, Germany}
\date{Date: \today}
\maketitle
\begin{abstract}
A general algorithm toward the solution of 
the fermion sign problem in finite-temperature
quantum Monte Carlo simulations has been formulated for 
discretized fermion path integrals with 
nearest-neighbor interactions in the Trotter direction.
This multilevel approach systematically implements a simple
blocking strategy in a recursive manner to 
synthesize the sign cancellations among different fermionic paths throughout
the whole configuration space. The practical usefulness
of the method is demonstrated for interacting electrons in a quantum dot.
\end{abstract}
\pacs{PACS numbers: 02.70.Lq, 05.30.Fk, 73.20.Dx}

\narrowtext

The quantum Monte Carlo (QMC) technique is one of the most powerful
methods for the simulation of many-fermion systems.  It is based on 
a path integral formulation of the fermion propagator and 
is one of the very few methods capable of delivering exact results for 
strongly correlated systems.
Despite its potentials, applications of QMC have been severely 
handicapped by the notorious ``fermion sign problem'' \cite{qmcgeneral,loh}.  
As a consequence of exchange,
fermionic density matrix elements are not positive-definite. 
The sign cancellations arising from sampling fermion paths then manifest
themselves as a small signal-to-noise ratio that vanishes exponentially
with either the system size or with decreasing temperature. 
Besides variational or approximate treatments such as the fixed-node 
approximation \cite{ceperley}, the sign problem has remained unsolved.

In this Letter, we propose a simple and intuitive
approach toward the general solution of the fermion
sign problem. Our algorithm represents the systematic
implementation of a {\em blocking strategy} \cite{own}.
The idea behind the blocking strategy is that by sampling
groups of states, the sign problem can always be reduced compared to 
sampling single states.  By suitably bunching states together
into blocks, the sign cancellations among states within the same block 
can be accounted for non-stochastically.  It can then be shown \cite{own}
that {\em any} such blocking will always reduce the sign
 problem --- no blocking
will ever make the sign problem more severe \cite{footnote1}.
Any real progress on the sign problem will require 
an accurate treatment of the sign cancellations 
within suitably chosen subunits (blocks) of state space.

A systematic improvement of the sign problem can be achieved 
by formulating the blocking strategy in a recursive bottom-to-top
fashion. Blocks of different sizes are defined on several
{\em levels}, and after taking care of the sign cancellations
within all blocks on a given (finer) level, the resulting sign 
problem can be transferred to the next (coarser) level.  By doing this
recursively, the sign problem on all the coarser levels can be handled 
in the same manner.  It is then possible to proceed without 
numerical instabilities from the bottom up to the top level, 
where the last remaining cancellations pose no serious challenge.

In many ways, the algorithm we are proposing is related to the renormalization
group approach.  But instead of integrating out information
on fine levels, the sign cancellations are ``synthesized''
within a given level and subsequently their effects are 
transferred to coarser levels. Our approach is actually closer in spirit to 
the multi-grid algorithm \cite{mg}.
The method of Ref.~\cite{own} can be understood as a uni-level scheme
which provides a partial solution to the sign problem.
In contrast, given sufficient computer memory,
the algorithm proposed here can provide a complete solution.
Below we describe this {\em multilevel blocking} (MLB) algorithm
and apply it to the simulation of correlated 
electrons in a quantum dot.

We consider a many-fermion system whose state is described by a set of
quantum numbers $\bbox{r}$ denoting, e.g., the positions and spins 
of all particles. For simplicity, we focus
on calculating the expectation value of a diagonal operator \cite{f1},
\begin{equation}\label{start}
\langle A\rangle = \frac{\sum_{\bbox{r}} A(\bbox{r})
\rho(\bbox{r},\bbox{r})}{\sum_{\bbox{r}} 
\rho(\bbox{r},\bbox{r})} \;,
\end{equation}
where $\sum_{\bbox{r}}$ represents either a summation for the case of a 
discrete system or an integration for a continuous system.
Imaginary time is then discretized into $P$ slices of 
length $\epsilon=\beta/P$, where $\beta = 1/k_B T$
and we require $P=2^L$. Inserting complete sets at each slice $m=1,\ldots,P$,  
and denoting the corresponding configuration on slice $m$ 
by $\bbox{r}_m$, the diagonal elements of the density matrix at
$\bbox{r}=\bbox{r}_P$ read
\begin{equation}\label{rho}
\rho(P,P) = \sum_{1,\ldots, P-1}
(1,2)_0 (2,3)_0 \cdots (P,1)_0 \;.
\end{equation}
As a shorthand notation, we use the slice index $m$ 
for the quantum numbers $\bbox{r}_m$.
This equation also defines the {\em level-0 bonds},
which are simply the short-time propagators,
\begin{equation}\label{short}
(m,m+1)_0 = \langle \bbox{r}_{m+1}| e^{-\epsilon H}
|\bbox{r}_m \rangle \;.
\end{equation}
This formulation of the problem excludes 
effective actions such as those arising from an integration over the fermions
via the Hubbard-Stratonovich transformation \cite{qmcgeneral,loh},
since they generally lead to long-ranged imaginary-time interactions.

\begin{figure}
\epsfxsize=0.8\columnwidth
\epsffile{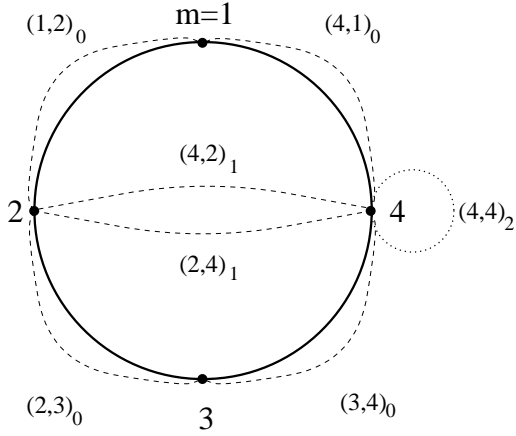}
\caption[]{\label{fig1}
Levels for $L=2$ ($P=4$). Imaginary time flows along the 
circle (solid curve), and the slices $m=1,2,3,4$ are distributed among
the three levels: The finest level $\ell=0$
contains  $m=1,3$, level $\ell=1$ contains $m=2$, and
 $\ell=2$ contains $m=4$. Level-$\ell$
bonds are indicated by dashed and dotted lines.
}
\end{figure}

To describe the MLB strategy, we need to specify the 
different levels $0\leq \ell \leq L$, where $L$
defines the Trotter number $P=2^L$.
Each slice $m$ belongs to a unique level $\ell$, such that $m=(2j+1)2^\ell$
and $j$ is a nonnegative integer.  For instance, the slices
$m=1,3,5,\cdots, P-1$ belong
to $\ell=0$,  $m=2,6,10,\cdots, P-2$ belong to $\ell=1$,
etc., such that there are ${\cal N}_\ell=  2^{L-\ell-1}$ 
(but ${\cal N}_L=1$)
different slices on level $\ell$, see Figure \ref{fig1}.
An elementary blocking is achieved by grouping together
configurations that differ only at slice $m$, so 
only $\bbox{r}_m$ varies in that block while all
$\bbox{r}_{m^\prime\neq m}$ remain fixed.
Sampling on level $\ell$ therefore 
extends over configurations $\{\bbox{r}_m\}$ living on the 
${\cal N}_\ell$ different slices.
In the MLB scheme, we move recursively from  
the finest ($\ell=0$) up to the coarsest level ($\ell=L$), and
the measurement of the diagonal operator is done only at the
top level using the configuration $\bbox{r}_P$.

We now describe a practical implementation of the MLB scheme.
A Monte Carlo sweep starts by changing only configurations
associated with the slices on level $\ell =0$ according to the weight
\begin{equation}
{\cal P}_0 = |(1,2)_0 (2,3)_0 \cdots (P,1)_0| \;,
\end{equation}
generating a MC trajectory containing $K$ samples for each slice 
on level $\ell=0$.  These ${\cal N}_0 K$ samples are stored and they are 
used to generate additional coarser interactions among the higher-level slices,
\begin{eqnarray}\nonumber
(m,m+2)_1 &=&  \langle {\rm sgn} [ (m,m+1)_0 (m+1,m+2)_0 ]
 \rangle_{{\cal P}_0[m+1]} \\ \label{l1b}
&=& ({\cal N}_0 K)^{-1} \sum_{m+1} {\rm sgn} [ (m,m+1)_0 \\ \nonumber 
&& \times\,   (m+1,m+2)_0] \;,
\end{eqnarray}
where the summation $\sum_{m+1}$ extends over the ${\cal N}_0 K$ samples.
As will be discussed in detail later on,
for a complete solution of the sign problem, 
the sample number $K$ should be chosen as large as possible.
The {\em level-1 bonds} (\ref{l1b}) contain crucial information about the
sign cancellations on the previous level $\ell=0$.
Using these bonds, the density matrix (\ref{rho}) is rewritten as
\begin{eqnarray} \nonumber
 \rho(P,P)&=&\sum_{1,2,\ldots,P-1}
 |(1,2)_0 (2,3)_0 \cdots (P,1)_0| \\ &\times&
 (2,4)_1 \cdots (P-2,P)_1 (P,2)_1 \;.
\end{eqnarray}
Comparing this to Eq.~(\ref{rho}), we notice that 
the entire sign problem has been transferred to the next coarser level 
by using the level-1 bonds. 

In the next step, the sampling is carried
out on level $\ell=1$ in order to generate the next-level bonds, i.e.,
 only slices $m= 2, 6,\ldots, P-2$ are updated,
using the weight ${\cal P}_0 {\cal P}_1$ with 
\begin{equation}
{\cal P}_1 = |(2,4)_1 (4,6)_1 \cdots (P,2)_1| \;.
\end{equation}
Moving the level-1 configurations modifies the level-0 bonds, 
which in turn requires that the level-1 bonds be updated.
A direct re-calculation of these bonds according to Eq.~(\ref{l1b}) 
would be too costly. Instead, we use the stored configurations 
on level $\ell=0$ to perform an importance sampling of the new level-1 bonds.
Under the test move $m \to m^\prime$ (i.e., $\bbox{r}_m\to \bbox{r}_m^\prime$)
on level $\ell=1$, the bond (\ref{l1b}) can be obtained from 
\begin{equation} \label{eq8}
(m^\prime,m+2)_1 = \frac{\sum_{m+1} \frac{(m^\prime,m+1)_0 (m+1,m+2)_0}
{|(m,m+1)_0 (m+1,m+2)_0|} } {\sum_{m+1} \frac{|(m^\prime,m+1)_0 
(m+1,m+2)_0|} {|(m,m+1)_0 (m+1,m+2)_0|} } \;,
\end{equation}
where $\sum_{m+1}$ runs over the previously
stored MC configurations $\bbox{r}_{m+1}$.
Note that for small values of $K$,
Eq.~(\ref{eq8}) is only approximative, and thus a sufficiently large
value of $K$ should be chosen.
With the aid of Eq.~(\ref{eq8}), we obtain the 
updated level-1 bonds with only moderate
computational effort. Generating a sequence of $K$
samples for each slice on level $\ell=1$, and storing these
${\cal N}_1 K$ samples, 
we then calculate the {\em level-2 bonds} 
in analogy to Eq.~(\ref{l1b}),
\begin{equation}\label{l2b}
(m,m+4)_2 =  \langle {\rm sgn}\left[ (m,m+2)_1
(m+2,m+4)_1 \right] \rangle_{{\cal P}_1 {\cal P}_0} \;,
\end{equation}
and iterate the process up to the top level $\ell=L$
using the obvious recursive generalization of Eqs.~(\ref{l1b})
and (\ref{l2b}) to define {\em level-$\ell$ bonds}.

Thereby the diagonal elements of the density matrix are obtained as
\begin{eqnarray}\label{rfinal}
\rho(P,P) &=& \sum_{1,2,\ldots,P-1}
 |(1,2)_0 (2,3)_0 \cdots (P,1)_0| \\ \nonumber &\times&
 |(2,4)_1 \cdots (P-2,P)_1 (P,2)_1| \\ \nonumber &\cdots&
 |(P/2,P)_{L-1} (P,P/2)_{L-1}| \, (P,P)_L \;.
\end{eqnarray}
By virtue of this algorithm, the sign problem is transferred step by step up
to the coarsest level. The expectation value
(\ref{start}) can thus be computed from 
\begin{equation}\label{expec}
\langle A \rangle = \frac{\langle A(P) \, {\rm sgn}(P,P)_L
\rangle_{\cal P}}{\langle{\rm sgn}(P,P)_L \rangle_{\cal P}} \;.
\end{equation}
The manifestly positive definite MC weight ${\cal P}$
used for the averaging in Eq.~(\ref{expec}) can 
be read off from Eq.~(\ref{rfinal}),
\begin{eqnarray}\label{mcwe}
{\cal P}&=&|(1,2)_0 (2,3)_0 \cdots (P,1)_0| \\ \nonumber &\times&
 |(2,4)_1 \cdots (P-2,P)_1 (P,2)_1| \\ \nonumber &\cdots&
 |(P/2,P)_{L-1} (P,P/2)_{L-1}| \, |(P,P)_L|  \;.
\end{eqnarray}
The denominator in Eq.~(\ref{expec}) gives the average sign
and indicates to what extent the sign problem has been solved. Under a naive 
application of the QMC technique, the average sign  decays
exponentially with $\beta$  and is typically close to zero. This
causes the numerical instabilities associated with the sign problem,
i.e., to obtain statistically relevant results requires exponentially long
CPU times. With the MLB algorithm, however, the sign problem can be 
completely eliminated.  The average sign remains close to unity for
any $\beta$, with a CPU time requirement that increases only
linearly. The price to pay for the stability of the
algorithm is the increased memory
requirement associated with having to store the sampled configurations
on the fine levels, which scales at worst quadratically in $K$.
 The example below demonstrates that 
modest memory requirements are sufficient to treat rather complex problems.

Next  we address questions concerning the {\em exactness} of the MLB 
approach for a  finite sample number $K$.  
Clearly, $K$ needs to be sufficiently large
to produce a reliable estimate for the level-$\ell$ bonds. 
If these bonds could be calculated exactly (corresponding
to the limit $K\to \infty$), the manipulations leading to 
Eq.~(\ref{rfinal}) yield the exact result. Hence for large
enough $K$, the MLB algorithm must (i) become exact 
and (ii) completely solve the sign problem. 
However, since the level-$\ell$ bonds can only be computed for finite $K$, 
the weight function ${\cal P}$ amounts to using a  noisy estimator,
which in turn can introduce bias into the algorithm \cite{kuti}.  
In principle, this problem could be avoided by using a linear
acceptance criterion \cite{kuti} instead of the algorithmically simpler
 Metropolis choice \cite{qmcgeneral}. 
But even with the Metropolis choice (which we used in the
example below), the bias can be
made arbitrarily small by increasing $K$.
Therefore, with sufficient computer memory, 
the MLB approach can be made to give numerically {\em exact}\, results.  
One might then worry about the actual value of $K$ 
required to obtain stable and exact results.  If this
value were to scale exponentially with $\beta$ and/or system
size, the sign problem would be present in disguise again.
Although we do not have a rigorous non-exponential bound on $K$,
our experience with the MLB algorithm indicates that
this scaling is at worst algebraic. 

We now illustrate the usefulness of the method for
$N$ interacting electrons confined in a {\em quantum dot} \cite{leo}.
Quantum dots are  two-dimensional artificial
atoms fabricated by means of suitable gates 
in semiconductor heterostructures. For simplicity, we 
consider only spinless electrons,
zero magnetic field, and a parabolic confinement potential.
We employ a symmetric Trotter breakup \cite{qmcgeneral}
for $H=H_1+H_2$,
\begin{equation}
H_1 = \sum_{j=1}^N \left( \frac{\bbox{p}_j^2}{2m^*}
+\frac{m^*\omega^2_0}{2}\bbox{x}_j^2 \right) \; , \quad
H_2 = \sum_{i<j}^N
\frac{e^2}{\kappa |\bbox{x}_i-\bbox{x}_j|} \;,
\end{equation}
where the positions (momenta) of the $N$ electrons
are $\bbox{x}_j$ ($\bbox{p}_j$), the dielectric constant
is $\kappa$, and $m^*$ is the effective mass.
The short-time propagator (\ref{short}) under $H_1$ is 
obtained by antisymmetrizing a product of $N$ 
harmonic oscillator propagators, leading to a 
fermion determinant. Since the determinant can change sign,
conventional QMC simulations run into the sign problem.
The MC updating then employs randomly chosen single particle moves 
on the momentary level $\ell$. This suffices for an 
efficient and ergodic sampling of configuration space.
Here we present results for the energy $E_N=\langle H \rangle$.
Since the direct evaluation of the kinetic energy would involve
a nonlocal operator, we have exploited the quantum 
virial theorem \cite{slater} in order to sample $E_N$. 
Simulations were done at $\hbar\beta\omega_0=6$ for two
different interaction strengths,
$l_0/a=2$ and $8$. Here  
$l_0=(\hbar/m^* \omega_0)^{1/2}$ is a confinement lengthscale, 
and $l_0/a=e^2/(\hbar \kappa l_0 \omega_0)$, 
with $a$ being the effective Bohr radius.
Trotter convergence was achieved at $L=3 \; (4)$ for $l_0/a=2 \; (8)$.
The $N=2$ exact diagonalization results of Ref.\cite{merkt} 
have been accurately reproduced, which also serves as
an independent check for our code.
The simulations have been carried out  on an IBM RISC6000/590 workstation.

\begin{table}
\caption{\label{table1}
MLB results for $N=8$ and $l_0/a=2$.
 $N_s$ is the number of samples (in $10^4$),
$t_{\rm CPU}$ the total CPU time (in hours),
 MB the required memory (in mega-bytes), and
$\langle {\rm sgn} \rangle$ the average sign.
Bracketed numbers are error estimates.}  
\begin{tabular}{llllll}\hline
$K$ & $N_s$ &  $t_{\rm CPU}$ & MB  &
$\langle {\rm sgn} \rangle$ & $E_N/\hbar\omega_0$ \\ \hline
1   & 120  &   95 &   2 & 0.02 & $48.6(3)$\\
100 &   7  &   33 &  14 & 0.48 & $48.43(8)$\\
200 &   9  &   83 &  30 & 0.63 & $48.55(7)$\\ 
400 &   8  &  174 &  64 & 0.73 & $48.53(9)$\\
600 &  10  &  308 &  96 & 0.77 & $48.54(8)$\\
800 &   9  &  429 & 150 & 0.81 & $48.59(8)$\\ 
\end{tabular}
\end{table}

To elucidate how the MLB algorithm works in practice, in
Table \ref{table1} we compare simulation results for $N=8$ electrons
at various values of $K$.  Compared to the naive approach ($K=1$), 
using a moderate $K=200$ already increases the average sign 
from $0.02$ to $0.63$, making it possible to
 obtain more accurate results from much fewer
samples.  The data in Table \ref{table1} also confirms that the bias 
can be systematically eliminated by increasing $K$, so that 
the energy found at $K\geq 200$ essentially represents the exact result 
(within error bars). 
As expected, the CPU time per sample scales only linearly with $K$,  
where the memory requirements grow at most quadratically with $K$. 

\begin{figure}
\epsfxsize=1.1\columnwidth
\epsffile{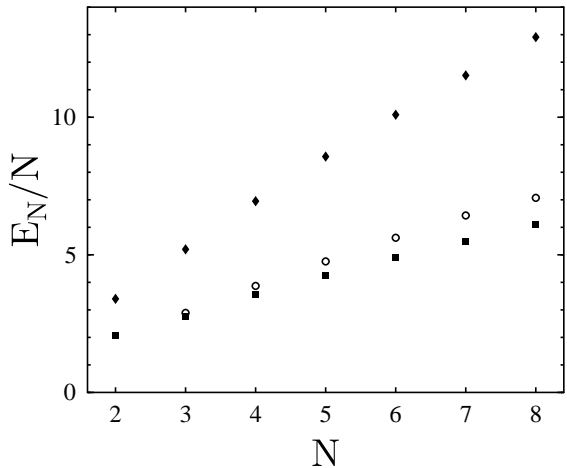}
\caption[]{\label{fig2}
Energy per electron, $E_N/N$,  in units
of $\hbar \omega_0$, for $l_0/a=2$ (squares) and $l_0/a=8$ (diamonds).
Statistical errors are smaller than the symbol size.
Open circles are taken from Ref.~\cite{bolton} for $l_0/a=2$.}
\end{figure}

Results for $E_N$ with $N\leq 8$ are shown in Figure \ref{fig2}.
For $N\leq 5$, the fixed-node QMC and Jastrow wavefunction calculations 
of Ref.~\cite{bolton} are in fairly good agreement with our exact results.
However, for larger $N$, there are deviations, 
with the correct energies significantly lower than the values 
reported in Ref.~\cite{bolton}, which represent variational upper bounds.
Notably, there are no obvious cusps or breaks in the
$N$-dependence of the energy.  Such features would 
hint at the existence of {\em magic numbers} for which the
artificial atom is exceptionally stable. Our data in 
Fig.~\ref{fig2} suggests that an explanation of the experimentally
observed magic numbers \cite{leo} has to involve spin or 
magnetic field effects.  Remarkably, the absence of 
pronounced cusps in $E_N/N$ for strong correlations ($l_0/a=8$) 
is in accordance with a  purely classical analysis \cite{classical}.

To conclude, we have proposed a multilevel blocking approach
to the fermion sign problem in finite-temperature QMC simulations.
As presented, the method applies to the primitive path integral
characterized by local imaginary-time interactions.
Given sufficient computer memory, the MLB
approach can provide a complete and exact solution of the sign problem. 
We believe that similar ideas may also lead to the resolution of the
sign problem in other fermion or real-time QMC schemes.

We thank H.~Grabert, W.~H\"ausler, and U.~Weiss for useful discussions.
This research has been supported by the National Science Foundation
under grants CHE-9257094 and CHE-9528121, by the Sloan Foundation,
the Dreyfus Foundation, and by the Sonderforschungs\-be\-reiche 
276 and 382 of the Deutsche Forschungsgemeinschaft (Bonn).


\begin{references}
\bibitem{f1}  By keeping two slices on the top level,
one could also sample nondiagonal operators.
\end{references}
\end{document}